\documentclass{iaus}
\usepackage{graphicx}

\title[Radial velocity follow-up] 
{Radial velocity follow-up for confirmation and characterization of transiting exoplanets}

\author[F. Bouchy et al.]   
{Fran\c{c}ois Bouchy$^1$, 
Claire Moutou$^2$, 
Didier Queloz$^3$, \\
\and the CoRoT Exoplanet Science Team 
}

\affiliation{
$^1$ Institut d'Astrophysique de Paris, UMR7095 CNRS, Universit\'e Pierre \& Marie Curie, \\
98bis Bd Arago, 75014 Paris, France, email: {\tt bouchy@iap.fr} \\[\affilskip]
$^2$ Laboratoire d'Astrophysique de Marseille (UMR 6110), 38 rue Frederic Joliot-Curie, \\
13388 Marseille Cedex 13, France \\[\affilskip]
$^3$ Observatoire de Gen\`eve, Universit\'e de Gen\`eve, 51 Ch. des Maillettes, 1290 Sauverny, 
Switzerland
}

\pubyear{2008}
\volume{253}  
\pagerange{1--10}
\jname{Transiting Planets}
\editors{F. Pont et al. eds.}
\begin{document}

\def\cms{\,cm\,s$^{-1}$}         
\def\ms{\,m\,s$^{-1}$}         
\def\kms{\,km\,s$^{-1}$}       
\def\vsini{$v$\,sin\,$i$}      
\def\Msol{M_\odot}             

\maketitle

\begin{abstract}
Radial Velocity follow-up is essential to establish or exclude the
planetary nature of a transiting companion as well as to accurately determine its
mass. Here we present some elements of an efficient Doppler 
follow-up strategy, based on high-resolution spectroscopy, 
devoted to the characterization of transiting candidates.  
Some aspects and results of the radial velocity follow-up of the CoRoT 
space mission are presented in order to illustrate the strategy used to deal with 
the zoo of transiting candidates.

\end{abstract}

\firstsection 
\section{Introduction}

The specific geometry of transiting planets makes them unique targets to obtain direct 
measurements of planetary radii and to give some insights on planet interiors 
and atmospheres. With the transit constraint on the system viewing angle, the mass 
of the transiting planet may be obtained by accurate radial velocity (RV) measurements and 
the planet mean density can be computed. The information on the planetary density is a 
strong observational constraints for theoretical 
studies on planets' structure and formation. Such parameters provide means to test 
scenarii of planet formation and evolution, for example to understand whether the planet possesses 
a core made of rocky material (\cite{guillot}). 
The special geometry of a transiting planet also permits interesting follow-up studies, 
such as searches for planetary satellites and studies of atmospheric features by 
transmission spectroscopy during the transit.

In the past few years, many extensive ground-based photometric surveys have been initiated 
to search for short-period transiting exoplanets, e.g. OGLE  
(\cite{udalski}), TrES (\cite{alonso04}), XO (\cite{mccullough}), 
HAT (\cite{bakos}) and SuperWASP (\cite{pollacco}). 
These surveys have yielded hundreds of planetary candidates. 
Out of this list, about 40 
transiting exoplanets\footnote{http://www.inscience.ch/transits/} have been 
established with an unambiguous determination of their mass thanks to the radial velocity follow-up. 
The numerous planetary candidates detected, compared to the quite 
small number of confirmed planets, explains the need to carry out 
intensive follow-up observations to complement transit surveys (\cite{bouchy05}) 
and illustrates the strong complementarity of both methods. 
In section 2, we discuss the different sources of confusion in planetary transit detection, 
illustrating the full necessity of radial velocity measurements. In section 3, we present 
some elements of the strategy of Doppler follow-up observations. 

Large ongoing ground-based transit searches are likely to find dozens of 
transiting Hot Jupiters, provided that the orbital period is short (P$\le$5 days) 
and the planetary radius is similar to Jupiter or larger.
Such ground-based surveys have access to the whole sky but are strongly limited 
by weather impacts, the diurnal cycle and the red noise limit of the Earth atmosphere. 
These effects prevent them from probing longer periods and smaller planets, leaving an 
unexplored domain of Neptune-like and rocky planets. These are the prime objective of the 
core planet-search program of the CoRoT mission. We present in section 4 some aspects and some results 
of the Doppler follow-up of CoRoT candidates.

\section{Impostors among transiting exoplanet candidates}

The detection of planetary transits suffers some ambiguity related to the 
system configuration, because different kinds of eclipsing systems may produce transit events 
that perfectly mimic planetary transits (\cite{brown}).
A candidate can be a transit of a low-mass M dwarf 
or brown-dwarf across a main sequence star, since in the low-mass regime the mass-radius 
relation is degenerate (eg: \cite{pont05}). 
It can also be a grazing eclipsing binary. Finally it can be an eclipsing binary (EB)
whose light is diluted by a third star. In the latter case, these might either be physical triple systems 
or a background eclipsing binary inside the instrumental PSF or photometric aperture. 
Follow-up observations, and in particular radial velocity measurements, are therefore 
mandatory to establish the planetary nature of a transiting companion. 
In other words, RV complementary observations are in all cases part of the detection process of 
transiting planets. Other approaches like deep analysis of light curve and transit 
shape, color analysis, high angular resolution photometry, and low resolution 
spectroscopy are not sufficient to ascertain the planetary nature of 
a transiting companion. High resolution spectroscopic measurements and radial velocity 
Doppler follow-up are the only ways to 1) ascertain the planetary nature, 2) characterize the 
properties of the central star, 3) characterize the true mass of the planet and finally 
4) to determine the sky-projected angle between the stellar spin and the planetary orbital axis 
(Rossiter-McLaughlin effect).

\section{Strategy for Doppler follow-up}

\subsection{Detailed light curve analysis}

Obviously, the first essential step is a careful analysis of the in- and out of transit light curve 
to select the transit events most likely originate from planets. 
The goal is to check the presence of secondary transit and/or ellipsoidal modulation in the light curve, 
the transit depth, shape and duration, as well as the color dependence if available. 
All these diagnostis may reveal eclipsing binaries.
It is also mandatory to precisely determine the significance of the transit detection 
with a correct estimation of noise including systematics (see \cite{pont06}). 
The radial velocity follow-up of false positives (wrong  detections) may be extremely 
telescope-time consuming. The threshold of transit detections should be carefully estimated taking 
into account all systematic effects and using a correct statistical estimator of the signal. 
Of course, all external available information should also be taken into account (color
information, spectral type, target contamination, ect ...). This initial step leads to 
a list of ranked candidates with confidence and planet-likelihood estimates.
  
\subsection{Identify binaries and/or multiple stellar systems}

An initial high-resolution spectroscopic measurement can provide information about 
the possible binarity (SB2) or multiplicity of the target. The Cross Correlation Function (CCF) 
of the spectrum may exhibit more than one component, indicating a SB2 or a multiple system. 

\subsection{Rule out giant stars with spectral classification}

A high-resolution spectrum with an intermediate signal-to-noise ratio permits to constrain 
the spectral type of the central star and to estimate its radius. 
This permits to rule out giant stars -- whose radius is not compatible 
with short-period planets -- and gives an estimate of the size of the transiting companion.

\subsection{Rule out blended-binary scenarios}

In some circumstances, the combination of a single star with a 
diluted unresolved EB can mimic both a planet transit signal and 
velocity variations. The velocity variations are then caused by 
the blend of both spectral components, the central star and the diluted background binary. 
In that case, the spectral lines of the fainter eclipsing binary move relative to the lines 
of the bright star and thus change the blended line-profiles. In order to 
examine the possibility that the radial velocity variation 
is due to a blend scenario, one needs to compute the CCF bisectors as described by 
\cite{santos} and check whether there is some correlation of the line 
asymmetries with phase. Furthermore, the CCF may be computed with different 
correlation templates in order to detect significant changes in the radial velocity values. Indeed, 
most blend scenarios produce template-dependent velocities due to the fact that the 
two blended components do not have the same spectral type.

\subsection{Radial velocity variations}

Light curve analysis provides the period $P$ and transiting epoch $T_0$ of 
the companion. Ideally, only 2 radial velocity measurements at $T_0 - P/4$ 
and $T_0 + P/4$ are sufficient to immediately determine the nature of the 
transiting body. 
Large radial velocity variations indicate a companion in the stellar mass range. 
For a circular orbit, the semi-amplitude of the radial velocity variation (in {\kms}) is 
related to the masses (in solar units) and period (in days) of the companion 
through:
\begin{equation}
K=214 \cdot \frac{m}{(m+M)^{2/3}} \cdot P^{-1/3}\,.
\end{equation}
A 0.1 solar-mass or 10 Jupiter-mass companion orbiting a solar-mass star 
with a 10-day orbit give respectively  $K$=9.3 and $K$=1.0 \kms. 
If the total amplitude of the RV variation is larger than a few \kms, the companion 
is clearly in the stellar mass range.

If the radial velocity variations are 
caused by a transiting object, then phase $\phi=0$ of the RV curve must correspond to the passage 
through center-of-mass velocity with a decreasing velocity.
If this is not the case, and the radial velocity variations do not concur with  
the period of the transit candidates, it is mandatory to re-analyze the light curve to 
check whether the transit signal is compatible with other periods. 
For instance, in a two-transit 
case on a time series with gaps, multiples of the interval 
between the two observed transits can lead to the wrong period. For equal-mass double-lined 
eclipsing binaries, the correct period is twice the candidate period, because both 
transits and anti-transits are present in the photometric curve.

If no significant RV variations are detected, it may indicate that the transit detection is a 
false positive, or that the eclipse does not come from the main target but from a diluted or 
background EB. Another alternative may be the lack of sufficient RV accuracy 
to detect the  Keplerian orbit of a planetary companion. These cases may be 
quite hard to solve and need other approaches to avoid wasting telescope time with useless 
Doppler observations. In such cases, the significance of the transit detection should be re-examined, 
high-resolution imaging of the candidate should be performed, the possible mass range of a 
planetary companion should be estimated with models, and finally an estimation of the
telescope time needed to more or less constrain the mass of the companion should 
be provided. 

If no peak is detected in the Cross Correlation Function, it may indicate either  that: 
1) the star is a fast-rotating early-type star (single or component of a multiple system); 
2) the target is synchronized by a short-period massive companion; or 3) the signal-to-noise is 
too low to obtain a reliable CCF.

\subsection{Measure the spectroscopic transit}

A measurement of the spectroscopic transit -- known as the Rossiter-McLaughlin (RM) effect -- 
defines the sky-projected angle between the planetary orbital axis and the stellar rotation axis. 
This effect may also be used to confirm the planetary size of the transiting body (from 
the amplitude of the RV anomaly) and its planetary mass (from the RV slope just outside the transit). 
In the case of a diluted eclipsing binary, the spectral lines of the fainter 
eclipsing binary move relative to the lines of the bright star and thus change the line profiles. 
In such a configuration, one should consider not only the flux ratio but the {\vsini}, the velocity zero 
point, and the spectral type of both systems. The probability of finding a configuration 
of blended eclipsing binary that could simultaneously reproduce the RV anomaly and 
the photometric light curve is quite small. Furthermore, if RVs are computed using different 
cross-correlation masks without a significant change in the shape and amplitude of the RM anomaly, one may 
consider that the spectroscopic transit confirms and secures the planetary nature of the transiting body. 

The amplitude of the signature $\Delta V_{\rm RM}$ is related to the {\vsini} and 
the radius ratio $r/R$ through the relation :
\begin{equation}
\Delta V_{\rm RM} \sim 1.1 \cdot (r/R)^2 \cdot vsini
\end{equation}

For stars with a high \vsini, the amplitude of the signature $\Delta V_{\rm RM}$ may be as large 
as that from the Keplerian orbit, and for long-period companions, the amplitude may even be greater. 
An advantage is that the signature of the spectroscopic
transit can be detected in a few hours (the transit duration) and does not require the 
observation along the whole orbital period. In that case, measurements are also less affected 
by stellar activity. However, activity-related RV jitter may locally change the apparent 
slope outside transit (eg: \cite {bouchy08}). Furthermore, one needs to sample the transit 
event with a sufficient number of measurements. This requires exposure times shorter than 1 hour 
with a photon-noise uncertainty smaller than the signature's amplitude.

\subsection{Exoplanet mass characterization}

When the transiting companion is clearly identified as a planet, 
the Doppler measurements are used to constrain its mass. 
The semi-amplitude $K$ is directly related to the mass of the planet, 
hence the uncertainty of the mass $\sigma_m$ is directly related to the uncertainty 
$\sigma_K$ of the parameters $K$. Considering a uniform distribution along the orbit 
of $N_{obs}$ RV measurements with an individual uncertainty $\sigma_{RV}$, the 
uncertainty $\sigma_K$ is given by :\\
\begin{equation}
\sigma_K= \frac{\sigma_{RV}}{\sqrt{N_{obs}/2}}\,.
\end{equation}

Note that in the particular case of RV measurements made only at the extrema of 
the Keplerian orbit, the relation becomes $\sigma_K=\sigma_{RV}/\sqrt{N_{obs}}$. 
One can estimate the number of RV measurements needed to constrain the mass at a given 
level following the relation:
\begin{equation}
N_{obs} = 2 \cdot \frac{\sigma_{RV}^2}{(K \cdot \frac{\sigma_m}{m})^2}\,.
\end{equation}
Table 1 presents some examples of $N_{obs}$, the number of independent measurements required, 
for different configurations. A radial velocity accuracy of 10 {\ms} allows the characterization 
of hot giant planets with a reasonable number of measurements, radial velocity accuracy of 1 {\ms} 
is mandatory to characterize hot telluric planets.

\begin{table}[h]
\begin{center}
\caption{Number of Doppler measurements at a given precision $\sigma_{RV}$ 
required to constrain the mass of a transiting planet with 10-day period as a function of 
the host star mass.}
{\scriptsize
  \begin{tabular}{lccc}\hline 
$\sigma_{RV}$=1 \ms     & $5\pm0.5 M_{Earth}$ & $10\pm1 M_{Earth}$ & $15\pm1.5 M_{Earth}$ \\
$\sigma_{RV}$=10 \ms    & $0.16\pm0.016 M_{Jup}$ & $0.31\pm0.031 M_{Jup}$ & $0.47\pm0.047M_{Jup}$ \\ \hline
 G2 (1.0 M$_{\odot}$) &  90 & 22 & 10  \\
 K0 (0.8 M$_{\odot}$) &  67 & 17 & 7  \\
 M0 (0.5 M$_{\odot}$) &  36 & 9 & 4 \\ \hline
\end{tabular}
}
\end{center}
\end{table}

\section{Doppler follow-up of COROT}

The CoRoT space telescope (\cite{baglin}), successfully launched in December 2006, 
is a low earth orbit 27-cm 
telescope with a wide-field camera (see also Baglin et al. and Barge et al., these proceedings). 
Each semester, observations consist of one 
long photometric run (LR) of 150 days and one short exploratory run (SR) of about 25 days. 
The sensitivity of CoRoT in the exoplanet channel permits the measurement of flux variations down to 
7.10$^{-4}$ (0.7 mmag) in one hour integration time on a V=15.5 star.
That level of accuracy is compatible with the signal of a solar-type 
star transited by a 2 $R_{\oplus}$ planet. During each run, both long and short, 
up to 12000 stars with V magnitudes in the range 12-16 are monitored with a sampling 
of 8.5 minutes.
As of June 2008, CoRoT has observed 5 fields: IRa01 (Feb - Apr 07), SRc01 (Apr - May 07), 
LRc01 (May - Oct 07), LRa01 (Oct 07 - Apr 08) and SRa01 (Apr - May 08). First analyses show that 
the instrument behaves extremely well and fulfills the expected specifications. The first light-curve 
processing required several months to be fine-tuned, and the reduced light curves were provided to the 
CoRoT community in December 2007. 
In the first short run IRa01, 26 planetary candidates were found. Subsequent ground-based follow-up 
was carried out mainly during the winter of 2007-2008 and allowed the confirmation and the characterization 
of 2 new transiting hot Jupiters (\cite{barge}, \cite{aigrain}, \cite{moutou}). 
Out of these 26 candidates, follow-up observations revealed, as expected, different kinds of eclipsing 
systems mimicking planetary transits: 7 cases of low-mass EBs, 2 cases of 
grazing EBs, 5 cases of background EBs (within the CoRoT aperture of 30 arcsec), 
and 5 cases of blended EBs. The nature of the 5 remaining candidates is not 
yet resolved. In the LRc01, 32 planetary candidates were found and preliminary ground-based 
follow-up conducted on 11 candidates during the summer of 2007 revealed 1 new transiting hot Jupiter 
and 1 brown dwarf (\cite{alonso08}; \cite{bouchy08}, \cite{deleuil}). The full analysis and 
intensive follow-up of LRc01 and SRc01 started in spring 2008 and additional results are expected. 
The LRa01 preliminary ground-based follow-up conducted on 6 candidates during the winter of 2007-2008 
revealed 1 new transiting hot Jupiter (Rauer et al., in prep). The full analysis and intensive 
follow-up of LRa01 and SRa01 will take place during the winter of 2008-2009. 

The experience of the follow-up of the IRa01 shows the strong need for intensive Doppler 
follow-up operations to perform the screening of candidates. Only about 10\% of the candidates 
are true transiting exoplanets. 
One particularly serious difficulty comes from faint and diluted background EBs, whose light 
merges in the CoRoT window with that of the primary target. Ground-based photometry allows the 
identification of such sources of confusing 
events, but is difficult due to strong scheduling constraints. Another approach 
consists in obtaining 1 or 2 spectroscopic measurements of the main contaminants within the 
photometric window of the target in order to check their binary nature, but such contaminants 
are faint targets (mv=15-20) for spectroscopy. 

Several high-precision spectrograph have already been involved 
in the Doppler follow-up of CoRoT planetary candidates, like FLAMES (8.2-m ESO), HARPS (3.6-m ESO), 
SOPHIE (1.9-m OHP), CORALIE (1.2-m La Silla) and the Coude Spectrograph (2.0-m TLS). 
These facilities are used in a complementary mode in order to observe all the candidates, identify their  
nature, and characterize the mass of the established planetary companions.
The following subsections illustrates some typical results obtained in the radial-velocity 
follow-up of CoRoT candidates.

\subsection{Eclipsing M-dwarfs}

\begin{figure}[!ht]
\centering
\includegraphics[angle=0,width=6cm]{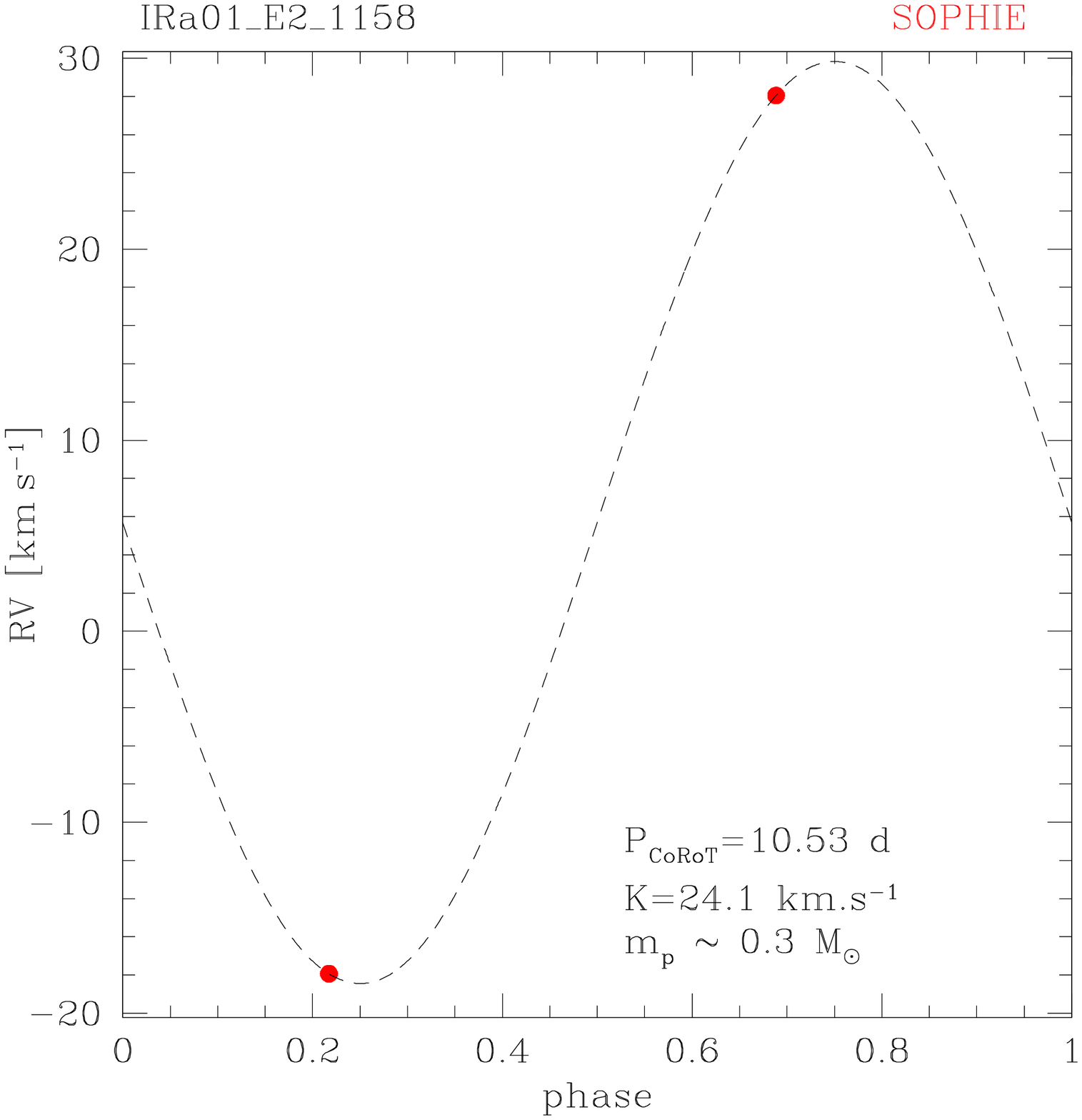}
\includegraphics[angle=0,width=6cm]{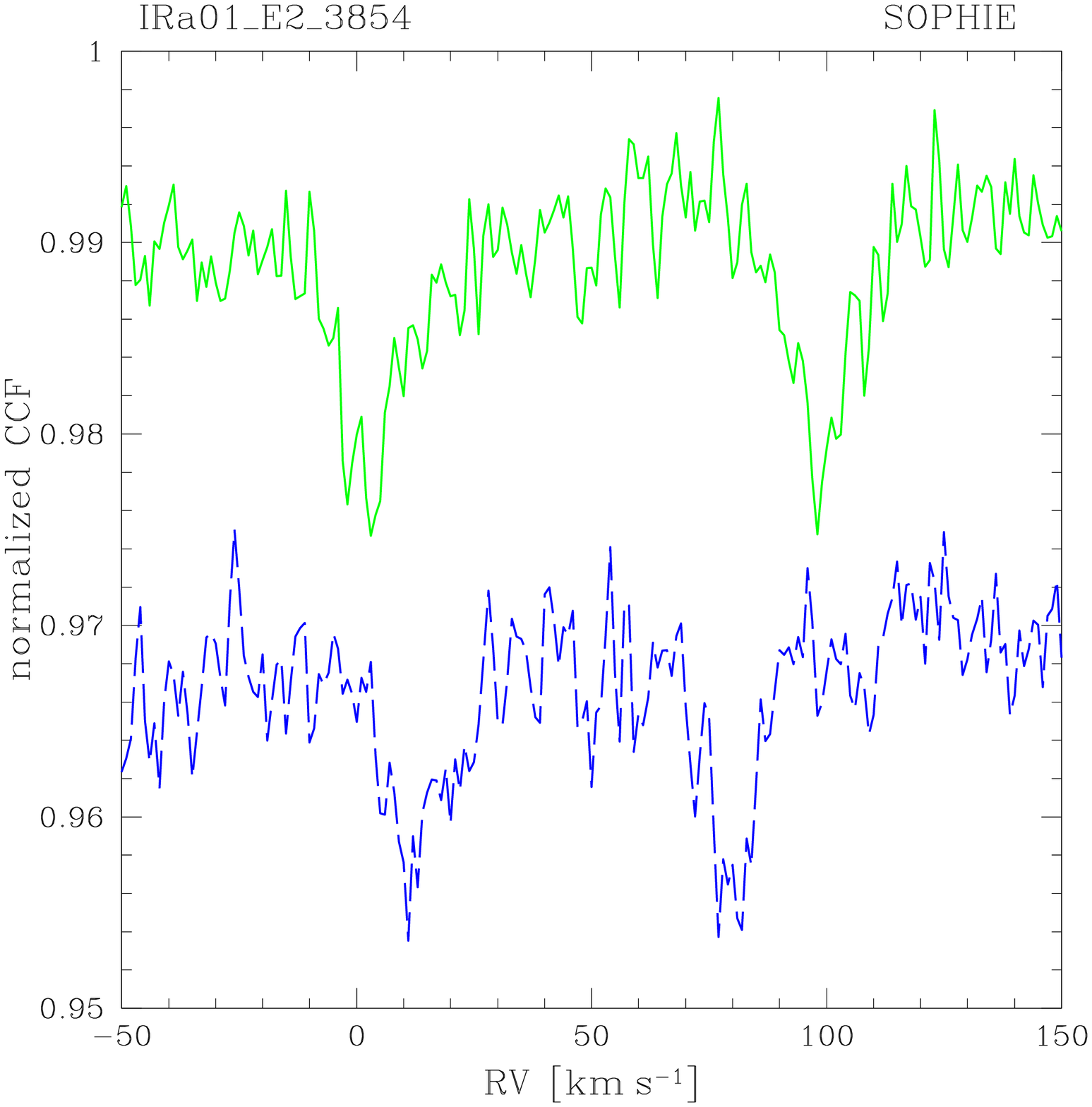}
\caption{(Left) Phase-folded RVs of the eclipsing M dwarf IRa01\_E2\_1158. (Right) Cross-Correlation 
Functions of the grazing eclipsing binary IRa01\_E2\_3854 observed at two different epochs.}
\label{fig1}
\end{figure}

A transiting low-mass star, like an M dwarf, is one of the easiest configuration and can be solved 
with only two radial velocity measurements. The radial velocity variation is typically larger than 
a few {\kms}. Figure \ref{fig1} (left) represents the phase-folded RV measurements made with SOPHIE on the 
CoRoT candidate IRa01\_E2\_1158 which shows a transit of 10 mmag on a V=13.2 star with a 
period of 10.5 days. 
The two measurements are in phase with the photometry and indicate that the transiting body is a
low-mass star of about 0.3 M$_\odot$. About 20\% of the CoRoT candidates correspond to this 
scenario.

\subsection{Grazing eclipsing binaries}

Grazing eclipsing binaries present two moving components in the Cross Correlation Function (CCF). 
One single RV measurement showing two components may reveal either a grazing EB or a 
unresolved triple system. A second RV measurement can indicate that the two components are moving 
in phase with the photometry thus revealing a grazing EB. In that case, the period found by 
RVs is twice the photometric period. Figure \ref{fig1} (Right) represents the CCFs of the CoRoT candidate 
IRa01\_E2\_3854 observed with SOPHIE at two epochs. This candidate shows a transit of 2 mmag on a 
V=15.3 star with a period of 1.14 days. About 10\% of the CoRoT candidates correspond to this 
scenario.

\subsection{Background eclipsing binaries (within CoRoT aperture)}

\begin{figure}[!ht]
\centering
\includegraphics[angle=0,width=7.5cm]{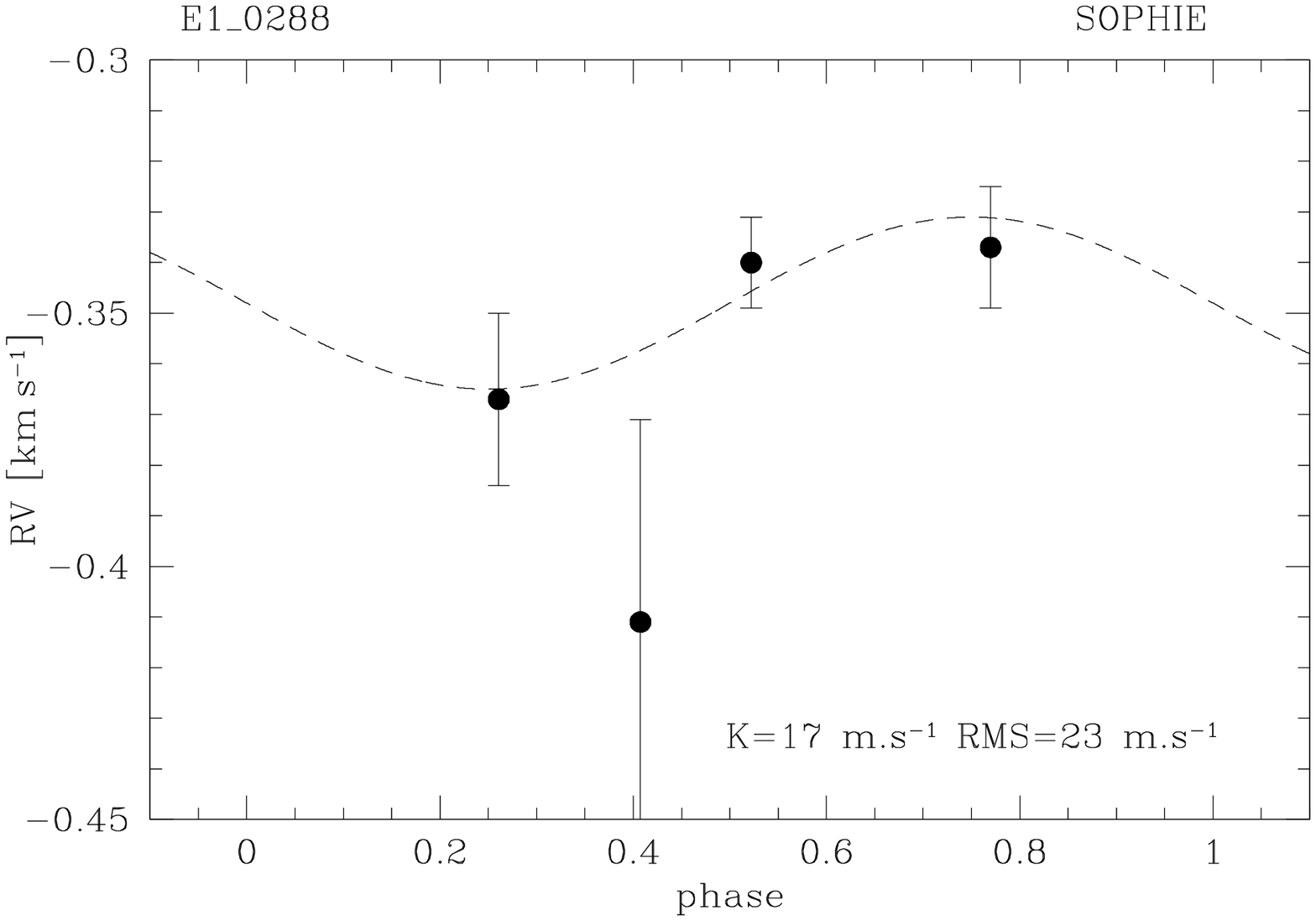}
\includegraphics[angle=0,width=5.5cm]{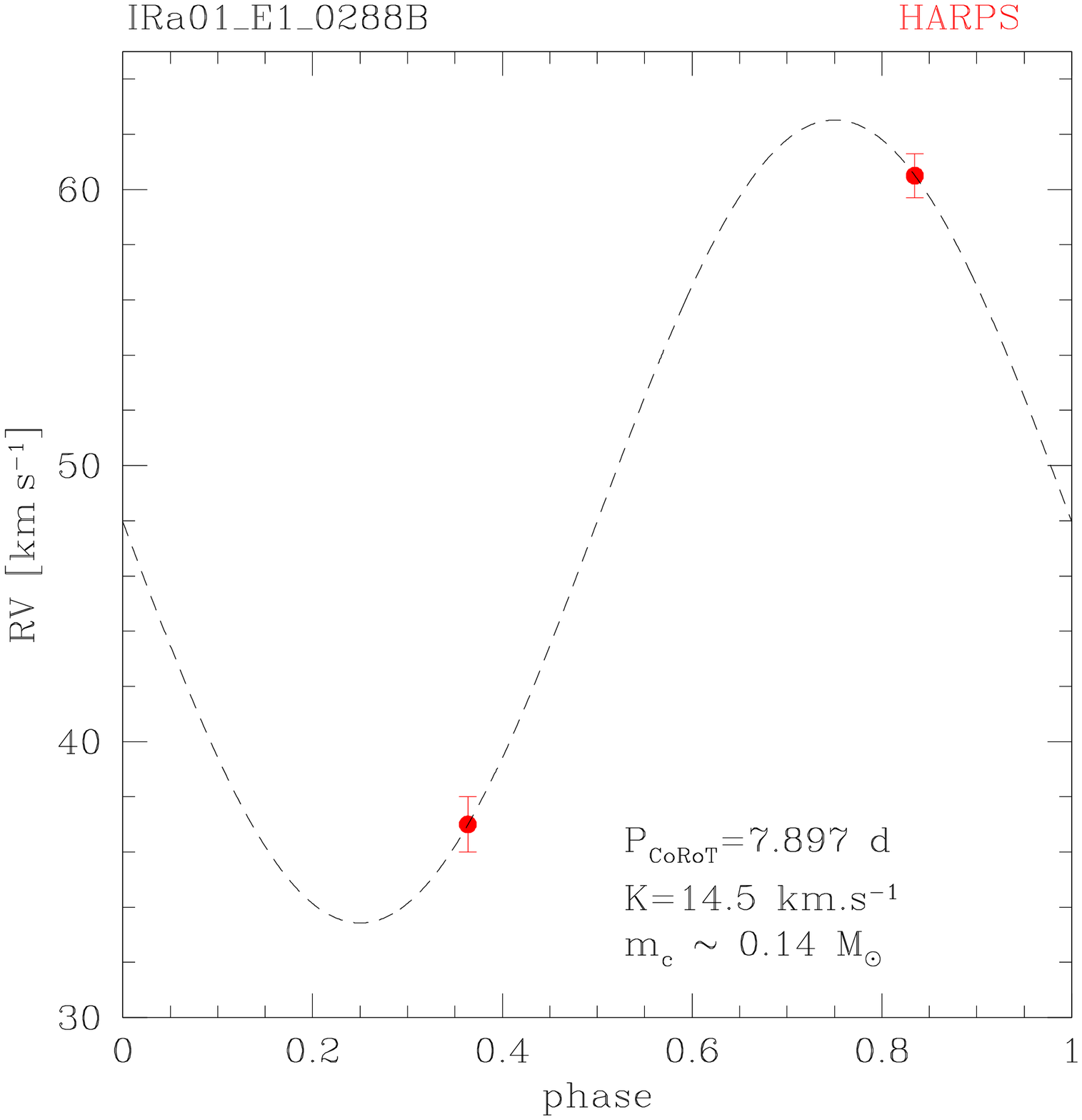}
\includegraphics[angle=0,width=6cm]{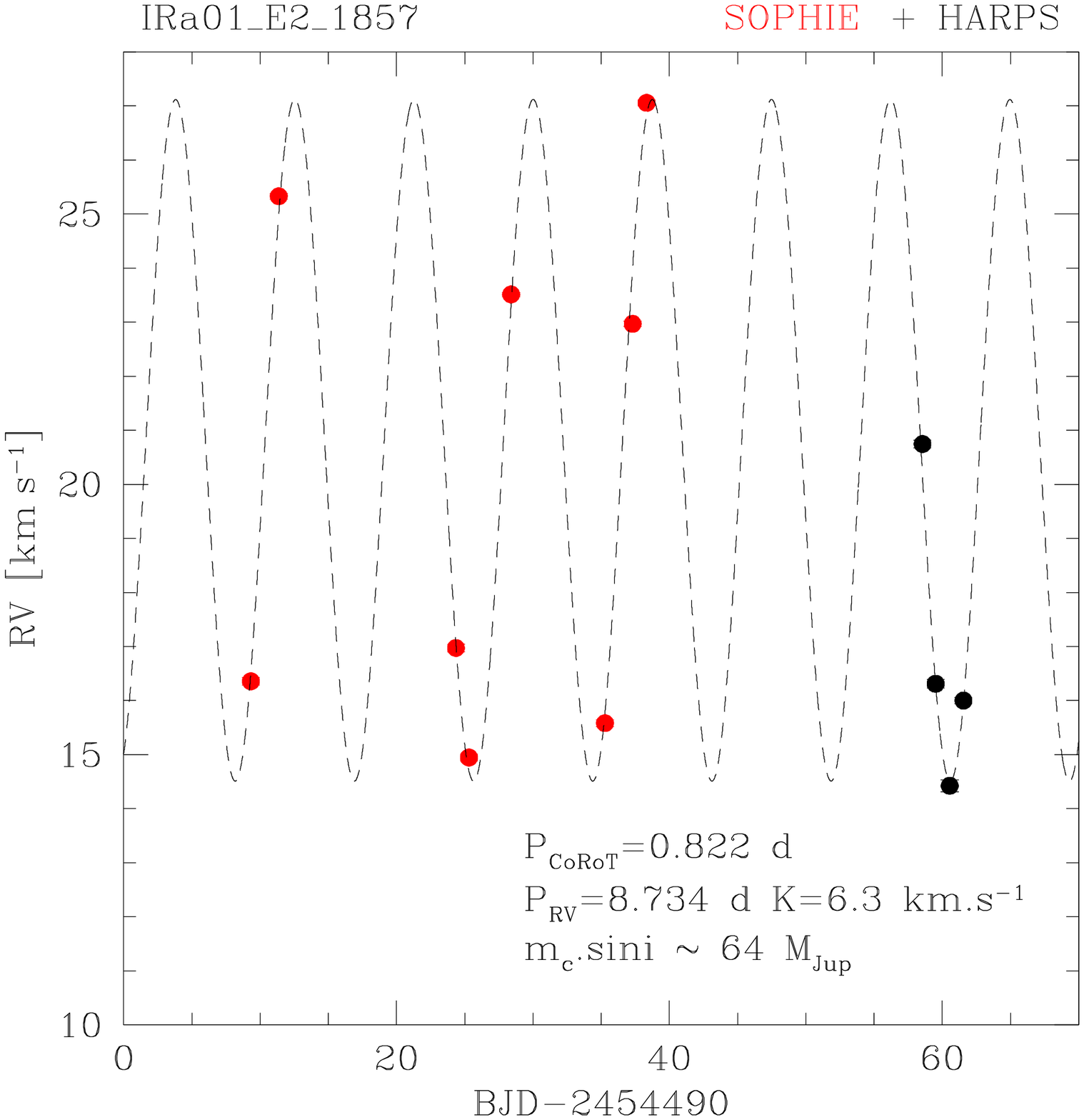}
\caption{(Top-Left) Phase-folded RVs of IRa01\_E1\_0288 observed with SOPHIE, indicating no significant 
RV variation. (Top-Right) Phase-folded RVs of the contaminant IRa01\_E1\_0288B observed with HARPS, 
indicating an eclipsing M dwarf.(Bottom) RVs of IRa01\_E2\_1857 indicating a low-mass stellar 
companion not related to the photometric signal.}
\label{fig2}
\end{figure}

The photometric apertures of CoRoT, with a diameter of 10 to 30 arcsec, typically includes several background contaminant stars. 
Some of them are EBs and their diluted signal inside the CoRoT window may mimic 
a planetary transit. Such a scenario cannot be solved with RV measurements. Indeed, if the EB contaminant 
is located several arcsec away from the main target, it is not observed by the spectrograph. 
In that case, ground-based photometry at higher spatial resolution than CoRoT is mandatory. 
However, in some cases RV measurements can be performed 
on the contaminant stars and reveal binaries. This is the case illustrated in Figure \ref{fig2} 
for the candidate IRa\_E1\_0288, showing a transit of 2.3 mmag on a V=13.5 star with a period 
a 7.9 days. SOPHIE RV measurements do not reveal a significant change. 
HARPS RV measurements made on the brighter contaminant star, a few arcsec away from the main target,
reveal the clear signature of a SB1. About 20\% of the CoRoT candidates correspond to this 
scenario.      

The case of the candidate IRa01\_E2\_1857 is another particular case of background eclipsing binary. 
It corresponds to a transit of 5 mmag on a V=14.3 star with a period of 0.82 days. Surprisingly, 
the RV measurements reveal a completely different period of 8.7 days (see Figure \ref{fig2} bottom). 
Additional ground-based photometry 
reveals that the transit is due to a background EB inside the CoRoT window. Independently of the CoRoT 
signal, the main target harbors a non-transiting companion with a minimum mass of about 64 Jupiter masses.

\subsection{Diluted and blended eclipsing binaries (within the seeing radius) }

In the case of a triple system or an unresolved background EB, the two components are observed 
by the spectrograph if they are closer than the seeing radius. 
If the two spectral components are blended, the spectral lines of the fainter 
eclipsing binary move relative to the lines of the bright star and thus change the blended 
line profiles, causing an apparent RV variation that is compatible with the signature of a planet. 
At least two approaches are used to solve such a scenario: 1) The CCF bisector variation 
may reveal such a blend; 2) if the two blended components do not have the same spectral type, 
the velocities will depend on the CCF template. 

\begin{figure}[!h]
\centering
\includegraphics[angle=0,width=6cm]{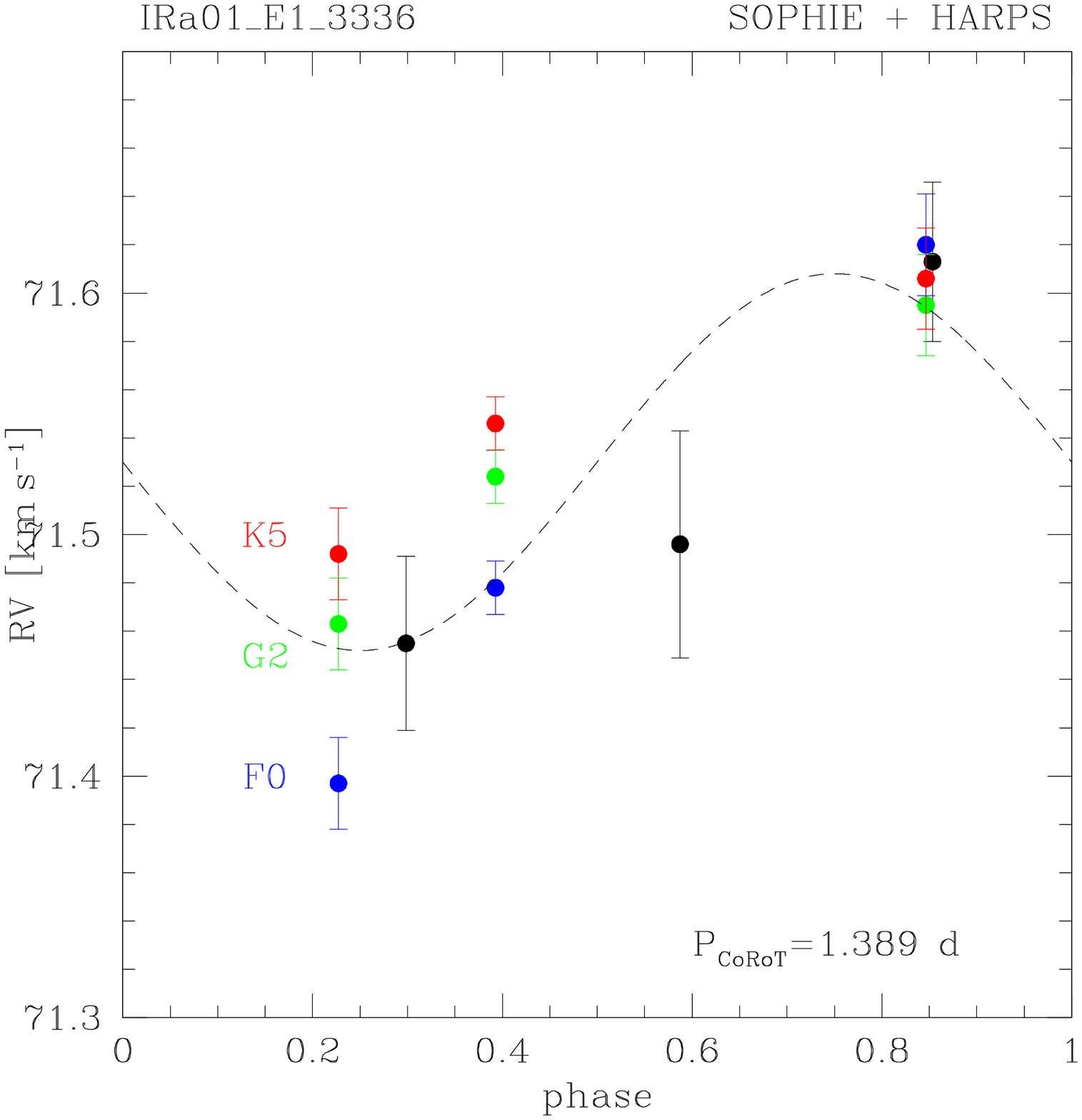}
\includegraphics[angle=0,width=6cm]{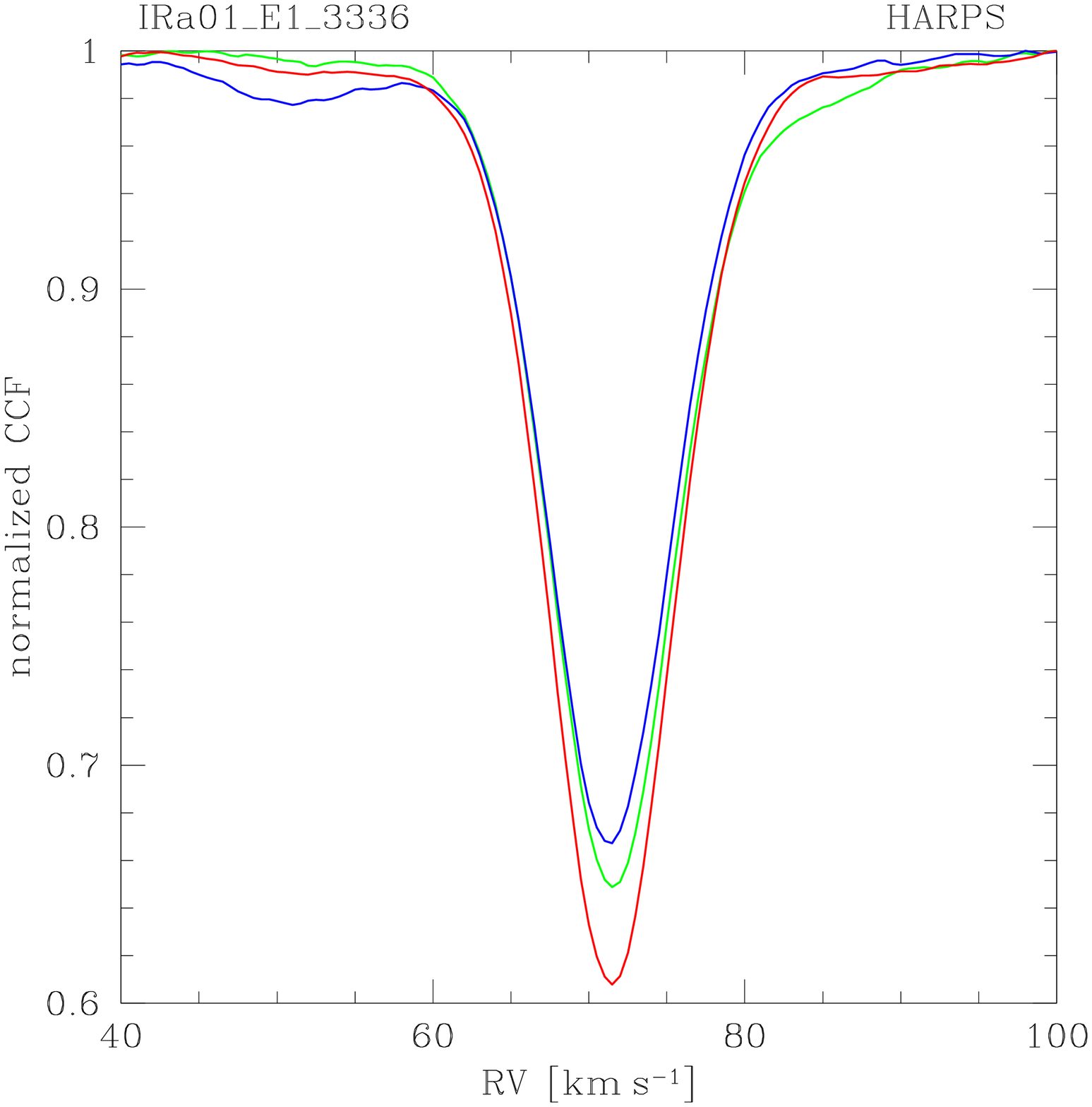}
\includegraphics[angle=0,width=6cm]{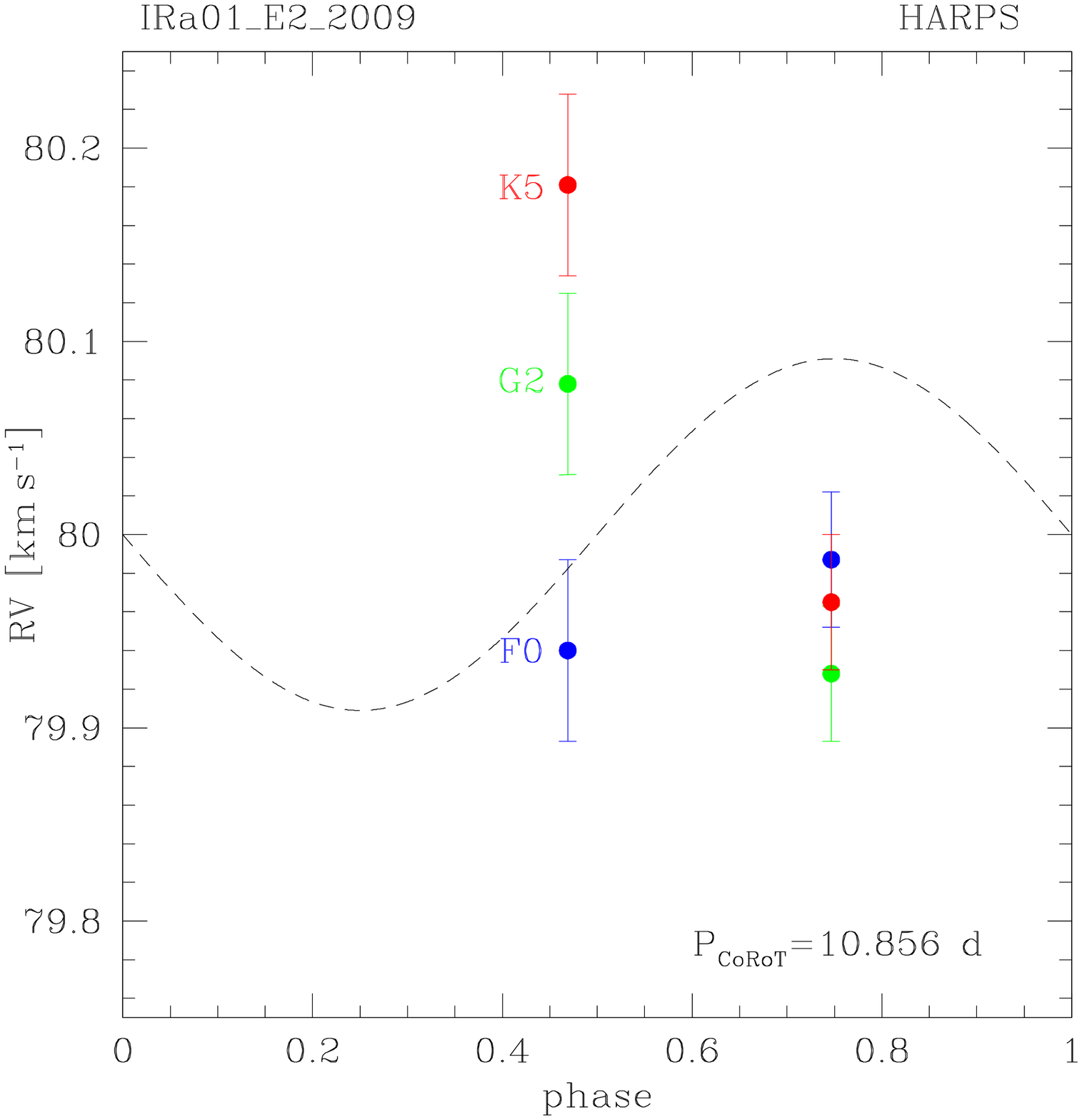}
\includegraphics[angle=0,width=6cm]{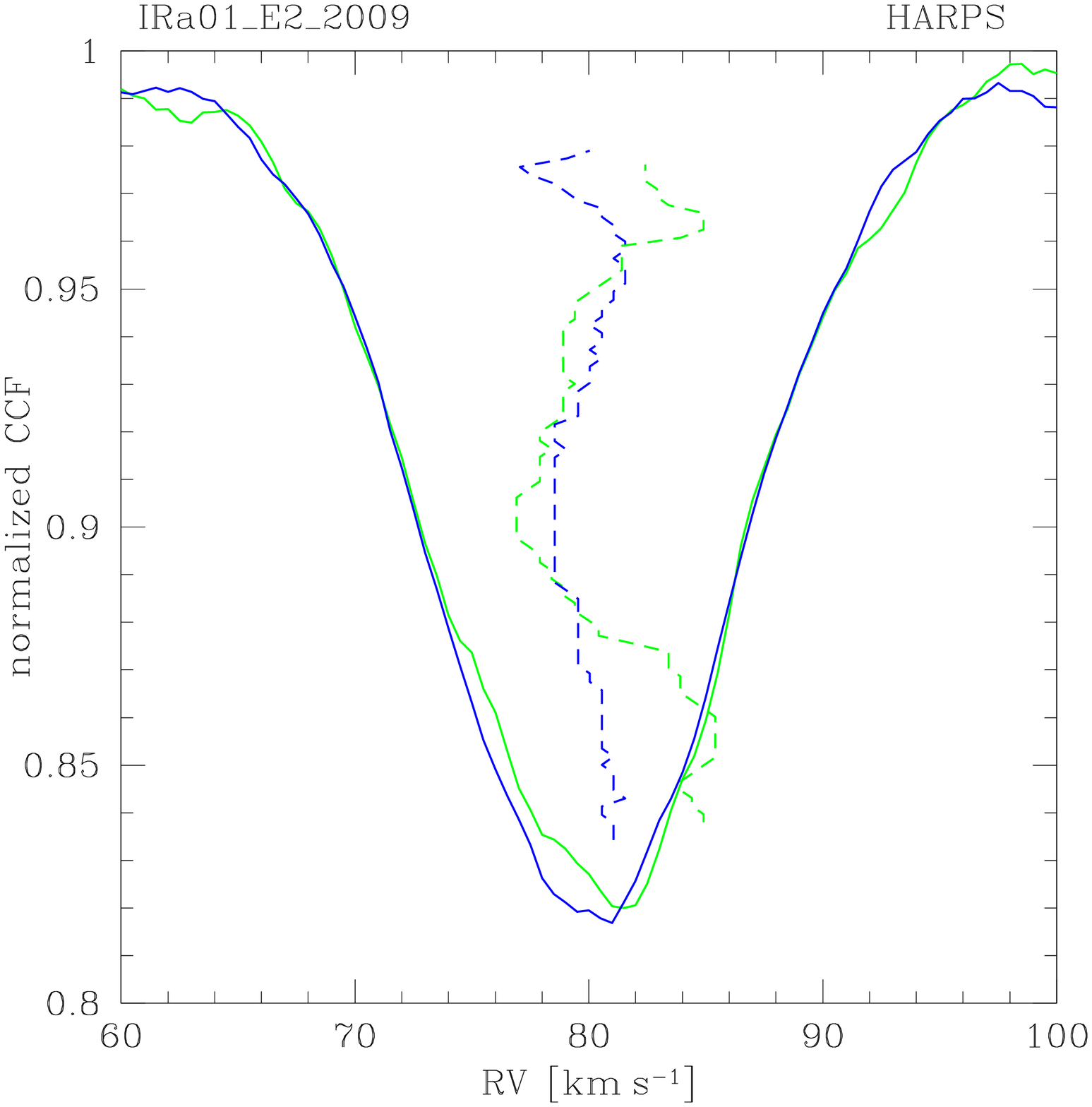}
\caption{Phase-folded RVs and Cross-Correlation Functions of the blended eclipsing 
binaries IRa01\_E1\_3336, when the spectra are correlated with masks of different star types (Top) 
and IRa01\_E2\_2009 (Bottom) with its bisector behaviour typical for a blended EB.}
\label{fig3}
\end{figure}

This is the case illustrated in Figure \ref{fig3} (top) with the candidate IRa01\_E1\_3336 
showing a transit of 1.7 mmag on a V=14.4 star with a period a 1.4 days. The RV amplitude changes 
with the CCF template and increases with the earlier type mask (F0), indicating that the background EB 
is an early-type star. A careful inspection of the CCF built with the F0 mask shows 
the second component moving across the main component.  
The candidate IRa01\_E2\_2009 (Figure \ref{fig3} bottom) also reveals a blended EB with 
a maximum amplitude obtained using the K5 CCF template. The bisector analysis reveals 
the blend. About 10\% of the CoRoT candidates correspond to this 
scenario.

\subsection{CoRoT Exoplanets}

\begin{figure}[!h]
\centering
\includegraphics[angle=0,width=9cm]{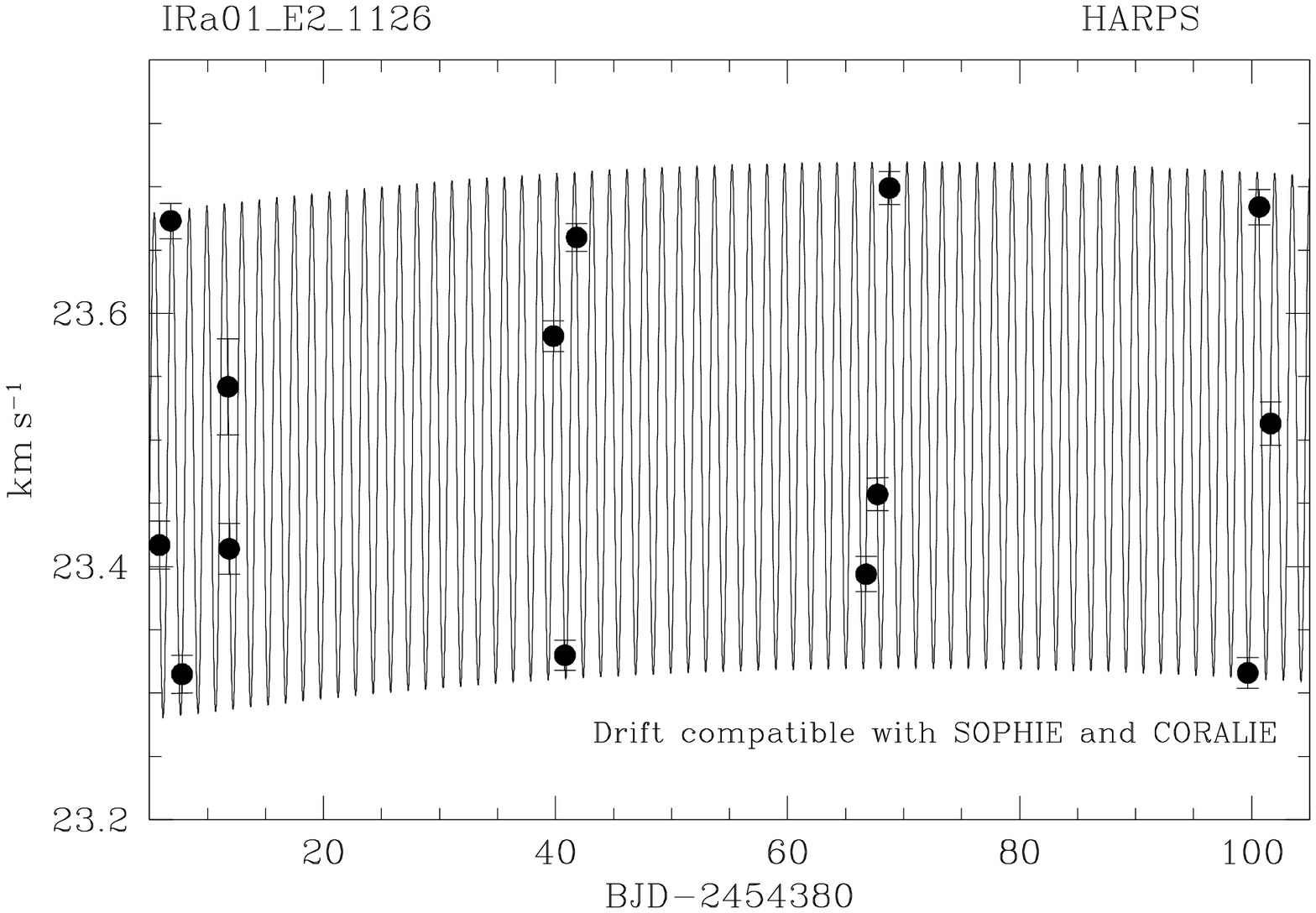}
\includegraphics[angle=0,width=9cm]{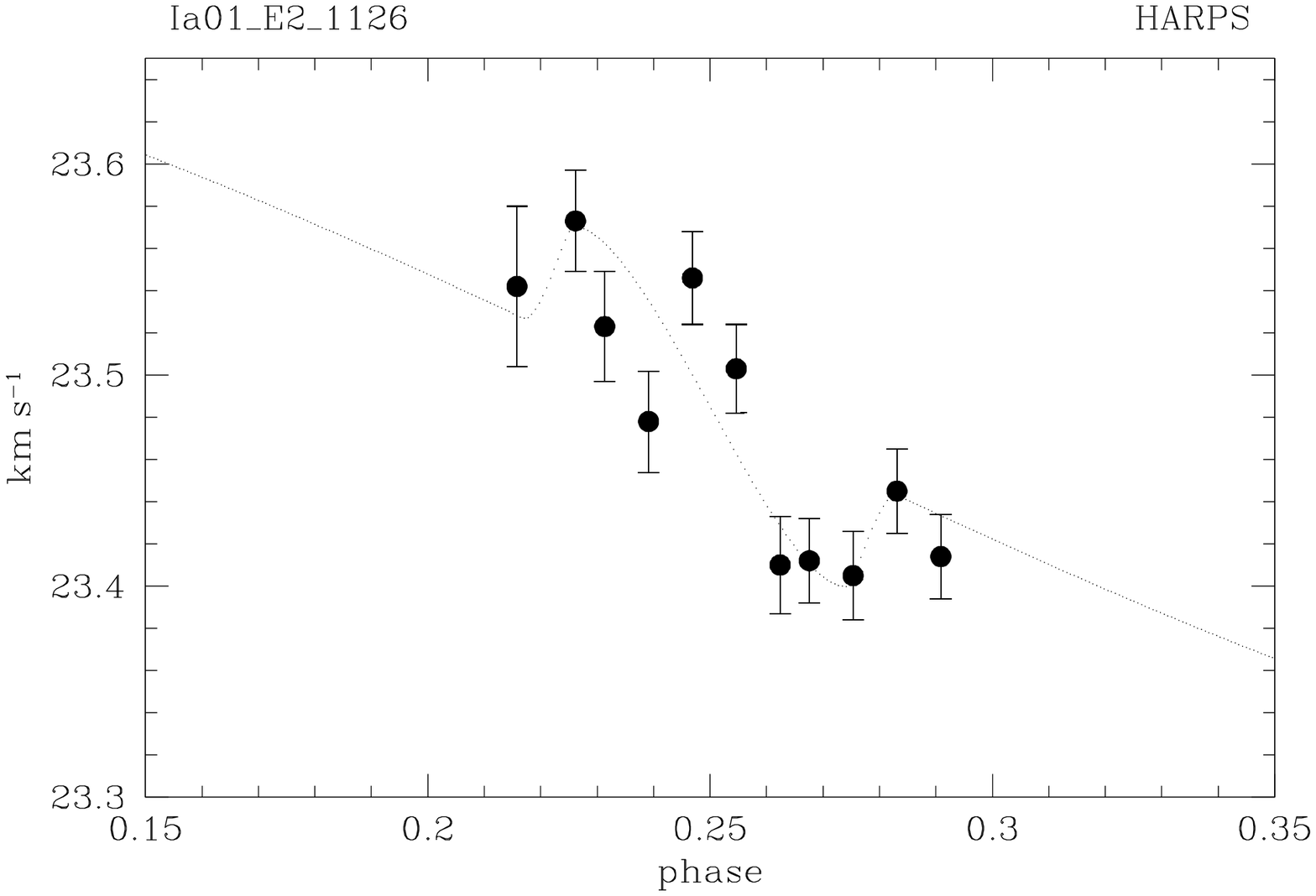}
\caption{(Left) RVs of CoRoT-exo-1 showing the 1.5-days Keplerian signal and a long-term drift 
due to a massive additional companion. (Right) RV anomaly (Rossiter-McLaughlin effect) of the 
spectroscopic transit of CoRoT-exo-1, as measured by HARPS in 2007.}
\label{fig4}
\end{figure}

As of June 2008, from the Doppler follow-up of 42 CoRoT candidates, 5 exoplanets have been identified 
and characterized. They are listed in Table 2. CoRoT-exo-1 (\cite{barge}) is still beeing monitored
with HARPS. A long term drift (see Figure \ref{fig4} top) reveals the presence of a long-period and massive 
second companion in the system. The Rossiter-McLaughlin effect was measured under quite bad conditions, 
but indicates that the planetary orbit is prograde with the stellar spin (see Figure \ref{fig4} bottom).       

\begin{table}[h]
\begin{center}
\caption{CoRoT exoplanets}
{\scriptsize
\begin{tabular}{lllll}\hline
Name     & Period & Mass & Radius & References \\
         & [days] & [M$_{\rm JUP}$] & [R$_{\rm JUP}$] & \\ \hline
CoRoT-exo-1b & 1.509 & 1.03 & 1.49 & \cite{barge} \\
CoRoT-exo-2b & 1.743 & 3.31 & 1.465 & \cite{alonso08}, \cite{bouchy08} \\
CoRoT-exo-3b & 4.26  & 21.6 & 0.97 & \cite{deleuil} \\
CoRoT-exo-4b & 9.202 & 0.72 & 1.19 & \cite{aigrain}, \cite{moutou} \\
CoRoT-exo-5b & 4.03  & 0.67 & 1.2  &  Rauer et al. in prep.\\ \hline
\end{tabular}
}
\end{center}
\end{table}

\section{Conclusion}

Radial Velocity follow-up is essential to confirm or exclude the
planetary nature of a transiting companion as well as to determine an
accurate mass. We have presented here some elements of an efficient Doppler 
follow-up strategy based on high-resolution spectroscopy 
devoted to the characterization of transiting candidates.  
Transit surveys illustrate the high need for RV follow-up to screen, solve 
and identify the true nature of transiting candidates. For example, from the OGLE 
2001 and 2002 seasons, 80 transiting candidates were followed and revealed 6 exoplanets. 
From the SuperWasp 2004 campaign, 47 transiting candidates were followed and revealed 
4 exoplanets. From the first year of operations of CoRoT, 42 transiting candidates were followed 
and revealed 5 exoplanets. Even from space, only about 10\% of candidates turn out to be 
true exoplanets. Radial velocity follow-up should be conceived as an integral part 
of these type of photometric surveys. In this context, one can appreciate the key role of small-class 
telescopes (from 1.2-m to 3.6-m), equipped with dedicated high-precision RV spectrographs, as they 
offer both flexibility and efficiency.

\end{document}